\documentclass[12pt]{article}
\usepackage{graphicx,epstopdf,tabularx,mathtools,amsfonts,bm,setspace,float,caption,subcaption}
\usepackage[pdfpagemode=UseNone,pdfstartview={XYZ null null null}]{hyperref}
\usepackage[longnamesfirst,numbers,square]{natbib}
\usepackage[margin=0.5in]{geometry} 
\usepackage{ulem}
\usepackage{latexsym}
\usepackage{amstext}
\usepackage{amsmath}
\usepackage{epsfig}
\usepackage{latexsym}
\usepackage{amstext}
\usepackage{amssymb}
\usepackage{graphics}
\usepackage{color}

\newcommand{\p}{\partial}

\newcommand{\beq}{\begin{eqnarray}}
\newcommand{\beqq}{\begin{eqnarray*}}
\newcommand{\eeq}{\end{eqnarray}}
\newcommand{\eeqq}{\end{eqnarray*}}
\newcommand{\eps}{\varepsilon}

\newcommand{\x}{\mbox{\boldmath$x$}}

\newcommand{\y}{\mbox{\boldmath$y$}}

\newcommand{\n}{\mbox{\boldmath$n$}}

\newcommand{\CC}{\mathcal{C}}

\newcommand{\FF}{\mathcal{F}}
\newcommand{\TT}{\mathcal{T}}

\definecolor{red}{rgb}{1,0,0}

\numberwithin{equation}{section}
\begin{document}
\title{Modeling ionic flow between small targets: insights from diffusion and electro-diffusion theory}
\author{Fr\'ed\'eric Paquin-Lefebvre$^{1}$ and David Holcman$^{1,\,2}$ \footnote{$^{1}$ Group of Data Modeling and Computational Biology, IBENS, \'Ecole Normale Sup\'erieure, 75005 Paris, France. $^2$ Department of Applied Mathematics and Theoretical Physics, and Churchill College, University of Cambridge, CB3 0DS, United Kingdom.}}
\date{\today}
\maketitle
\begin{abstract}
The flow of ions through permeable channels causes voltage drop in physiological nanodomains such as synapses, dendrites and dendritic spines, and other protrusions. How the voltage changes around channels in these nanodomains has remained poorly studied. We focus this book chapter on summarizing recent efforts in computing the steady-state current, voltage and ionic concentration distributions based on the Poisson-Nernst-Planck equations as a model of electro-diffusion. We first consider the spatial distribution of an uncharged particle density and derive asymptotic formulas for the concentration difference by solving the Laplace's equation with mixed boundary conditions. We study a constant particles injection rate modeled by a Neumann flux condition at a channel represented by a small boundary target, while the injected particles can exit at one or several narrow patches. We then discuss the case of two species (positive and negative charges) and take into account motions due to both concentration and electrochemical gradients. The voltage resulting from charge interactions is calculated by solving the Poisson's equation. We show how deep an influx diffusion propagates inside a nanodomain, for populations of both uncharged and charged particles. We estimate the concentration and voltage changes in relations with geometrical parameters and quantify the impact of membrane curvature.
\end{abstract}
\section{Introduction}
The flow of particles (molecules or ions) entering through narrow channels on the cellular membrane regulates the concentrations but also the potential equilibrium in excitable cells such as neurons \cite{hille,eisenberg}. While the selection of particles going through a channel and the amplitude of the particle current inside a single channel has been well characterized over the past 50 years, how entering particles evolve inside the cells in the microdomain around the single channel remains less explored. However, these regions can already contain neighboring channels that could be activated to respond as soon as possible to the first entrance. In particular, this could influence locally the membrane potential, modulated by the influx and efflux of ions across these neighboring channels \cite{bezanilla2000,bezanilla2008}. These local influxes modify constantly the local ionic concentrations and the electric field in small nanoregions, difficult to access experimentally. \\
In some cases involving small compartments such as cilia \cite{fain2019sensory}, dendritic spines \cite{yuste2010dendritic,holcman2015nrn} and glial protrusions, a  local difference of concentration generated by the influx through a channel could be much larger than the background ion concentration. For instance, calcium concentration inside a dendrite is of the order of a few 10 to 100 nM at rest \cite{Segalreview}, while calcium influx through channels can lead to a concentration ranging from 1 to 100 of $\mu{\rm M}$, a factor of at least 10. These concentration differences cause a local electric field and voltage drop that triggers the opening of neighboring channels or receptors \cite{zamponi2011} (Fig.~\ref{fig:channels}\textbf{C})). However the size of this region as well as the exact mechanism of ionic regulation in these regions near channel exits remain unclear.\\
This chapter summarizes recent efforts in modeling and computing the voltage and the dynamics of ions at the channel exit in microdomains containing other regulatory channels that influence the concentration of entering ions. The simplest modeling approach is to consider that ions are simply diffusing and then account for possible electrical charge interactions using an electro-diffusion model \cite{schussEisenberg2001,BenoitRoux}. Electro-diffusion models are described by the Poisson-Nernst-Planck equations, which consist in coupling the Fokker-Planck description for the density with the Poisson's equation for the voltage. One way to simplify the static charge case is to use classical Boltzmann solutions, for which ionic densities are exponential functions of the voltage, within the Poisson's equation, thereby resulting in a single nonlinear model equation. This approach, known as the Poisson-Boltzmann theory, is appropriate to describe properties of electrolytes near one or two planar membranes \cite{andelman,BenYaakov2009,orland2000} and can serve to introduce spatial scales, such as the Debye length, quantifying local charge interactions. However ionic densities given by Boltzmann solutions cannot account for sudden localized ionic fluxes resulting from channels and pumps dynamics, which is exactly the paradigm of our study: ions enter through one window, spread inside a microdomain (such as a ball, a cylinder, etc\ldots) and leave through another ensemble of target windows (Fig.~\ref{fig:fig1}\textbf{A})-\textbf{B})). \\
The chapter is organized as follows. In section \S \ref{sec:background} we introduce briefly the electro-diffusion modeling based on the Poisson-Nernst-Planck equations. In section \S \ref{sec:diffusion} we present, in the diffusion limit (neglecting electrochemical gradients), the computation of the concentration and the flow between two channels for a single species by solving the Laplace's equation. Then in section \S \ref{sec:electro_two} we consider two ionic species, positive and negative, that can interact. We compute the voltage profile using the Poisson-Nernst-Planck equations. We present some asymptotic analyses and the resulting biophysical formulas revealing the role of geometrical parameters on the voltage dynamics. Finally, we briefly conclude in section \S \ref{sec:conclusion}, summarizing the asymptotic and simulation difficulties and discuss some perspectives and implications in biophysics.
\section{Modeling ionic fluxes exiting a channel inside a domain }\label{sec:background}
\begin{figure}[!ht]
\centering
\includegraphics[width=0.75\linewidth]{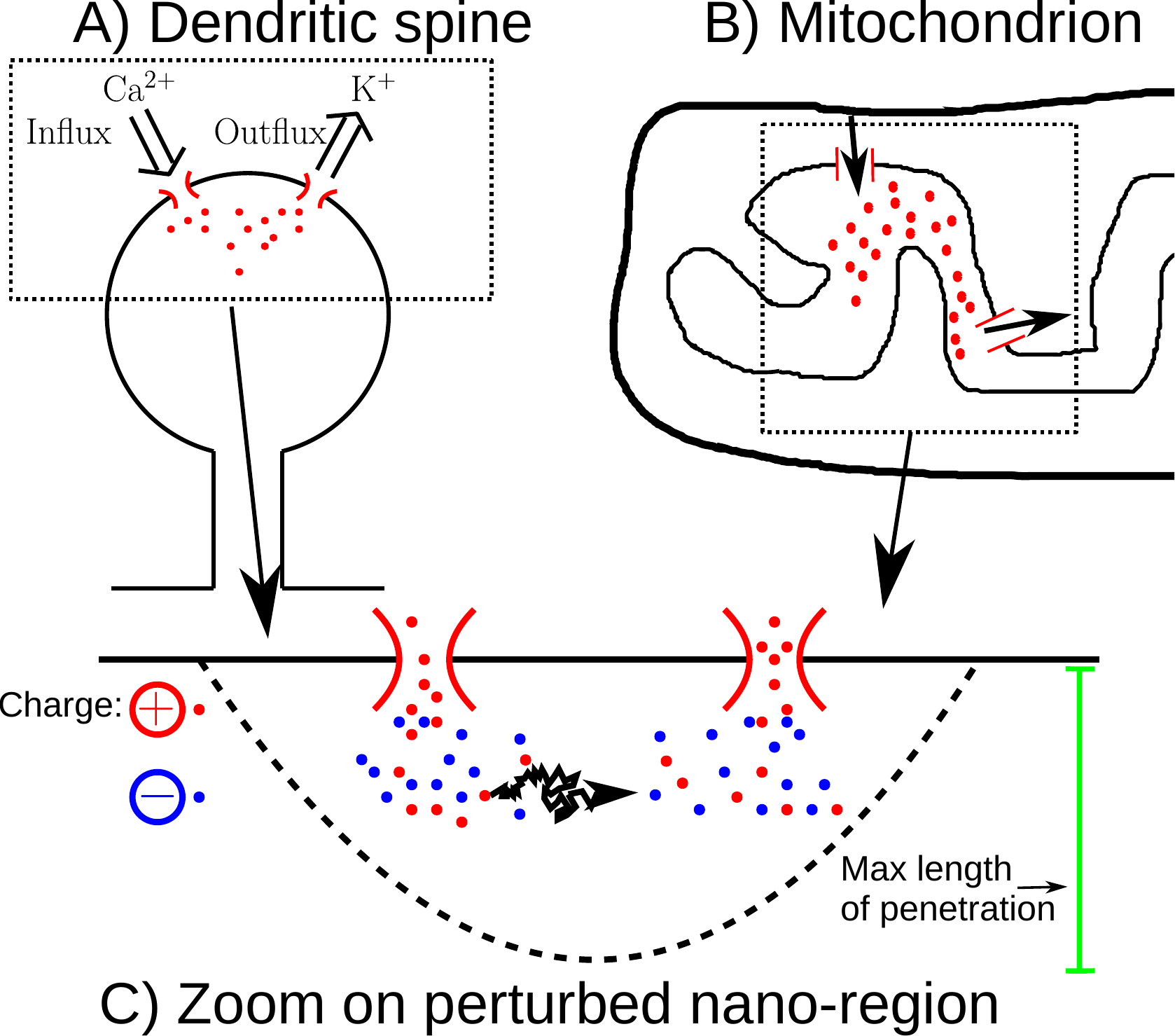}
\caption{\label{fig:channels} \textbf{Channels dynamics and voltage in neuronal nano-domains.} \textbf{A}) Dendritic spine receiving a calcium ions influx. \textbf{B}) Inner and outer membranes of a mitochondrion, with the inner membrane curvature thought to regulate voltage and calcium ions intake \cite{emboJ}. \textbf{C}) Formation of a voltage nano-domain in the vicinity of two ion channels (cf.~\cite{zamponi2011}).}
\end{figure}

Voltage changes in excitable cells such as neurons rely on currents generated by ionic channels opening and closing \cite{hille}. At the cellular level, voltage dynamics in neuronal cells is well characterized generating depolarization, bursts of activity or action potentials, as revealed by patch-clamp measurements or voltage sensitive dyes \cite{cartaillerNeuron}. However, voltage dynamics due to ionic transport in small micro-compartments (Fig.~\ref{fig:channels}\textbf{A})), is much less known due to spatio-temporal challenges, often below the diffraction limit, thus preventing a direct  experimental access.\\
In the absence of direct measurement, modeling and numerical simulations provide instructive insights on possible physiological mechanisms underlying voltage changes. By neglecting concentration changes voltage can be modeled by the classical cable theory \cite{tuckwell1988}. The one-dimensional voltage propagation within a neuron has provided a theoretical foundation for electrophysiology \cite{rall1962}. However this theory breaks down when applied to small neuronal compartments of the scale of $[0.1,\,1] \, \mu{\rm m}$, like dendritic spines, synaptic terminals or small neuronal processes such as filopodia or glial protrusions, because they assume spatial ionic homogeneity \cite{qian1989}. Indeed, voltage changes can be generated by a concentration gradient in neuronal nanodomains \cite{holcman2015nrn}. How voltage is regulated and spatially extends within such domains remain unclear: ionic concentration gradients may well be key elements, although they should be studied by further accounting for the local electric field generated by these gradients.\\
To study voltage changes at a subcellular level, the Poisson–Nernst–Planck (PNP) mean-field equations are used to model electro-diffusion, where diffusive ions can interact and generate an electric field. This description accounts for the geometry in computing the flow of electrical current. For a single charge density $c(\x,t)$ of electronic valence $z$ within a domain with boundary $\p\Omega$, the Nernst-Planck equation formulates as
\begin{equation}
\left\{
\begin{array}{ll}
\dfrac{\p c(\x,t)}{\p t} &= D\left( \Delta c(\x,t) + \dfrac{ze}{k_B \TT} \nabla (c(\x,t) \nabla v(\x,t)) \right)\,, \quad \x \in \Omega\,, \\& \\
0 &= D\left( \dfrac{\p c(\x,t)}{\p \n} + \dfrac{ze}{k_B\TT} c(\x,t)\dfrac{\p v(\x,t)}{\p\n} \right)\,, \quad \x \in \p\Omega \,,
\end{array}
\right.
\end{equation}
where $D$ is the diffusion coefficient and $k_B\TT/e$ is the thermal voltage. The voltage $v(\x,t)$ is solution of the Poisson's equation
\begin{equation}\label{eq:p1c}
\Delta v(\x,t) = - \frac{z\FF}{\eps\eps_0}c(\x,t)\,, \quad \x \in \Omega\,, \quad \frac{\p v(\x,t)}{\p\n} = - \sigma(\x,t)\,, \quad \x \in \p\Omega\,,
\end{equation}
where $\sigma(\x,t)$ is the surface charge density on the dielectric membrane. For such a class of non-electroneutral electro-diffusion models, charges accumulate near the boundary \cite{cartailler2017physD,cartailler2017scirep} and high membrane curvature as yielded by cusp or funnel-shaped boundaries plays a significant role in regulating internal voltage changes \cite{cartailler2017jns,cartailler2019jmb}. However to reduce voltage effects caused by large charge imbalances, more realistic electro-diffusion models should incorporate multiple charges, with global (but not necessarily local) electro-neutrality condition. Interestingly for an immobile excess of negative charges at the center of a ball domain \cite{tricot2021}, voltage effects were found to spatially extend on scales much longer than the Debye Length defined as \cite{andelman} $\lambda_D = \sqrt{\frac{\eps \eps_0 k_B\TT}{z^2e^2N_A2C_0}}$, which can be less than one nanometer for background ionic densities of hundredths of millimolars.\\
One goal of this book chapter is to present how non-local electroneutrality, with spatial charge imbalance coming from a localized flux of positive ions on narrow boundary patches, can affect the voltage inside the nanodomain (\S\ref{sec:electro_two}). This analysis could be applied to calcium or sodium ions entering neuronal nanocompartments via ionic channels, disturbing locally the voltage distribution and causing the formation of a voltage drop nanodomain, and finally leading to the activation of nearby potassium channels from which excess of positive charges will exit \cite{zamponi2011} (Fig.~\ref{fig:channels}\textbf{C})).
\section{Influx diffusion steady-state models for a single species}\label{sec:diffusion}
\begin{figure}[!ht]
\centering
\includegraphics[width=\linewidth]{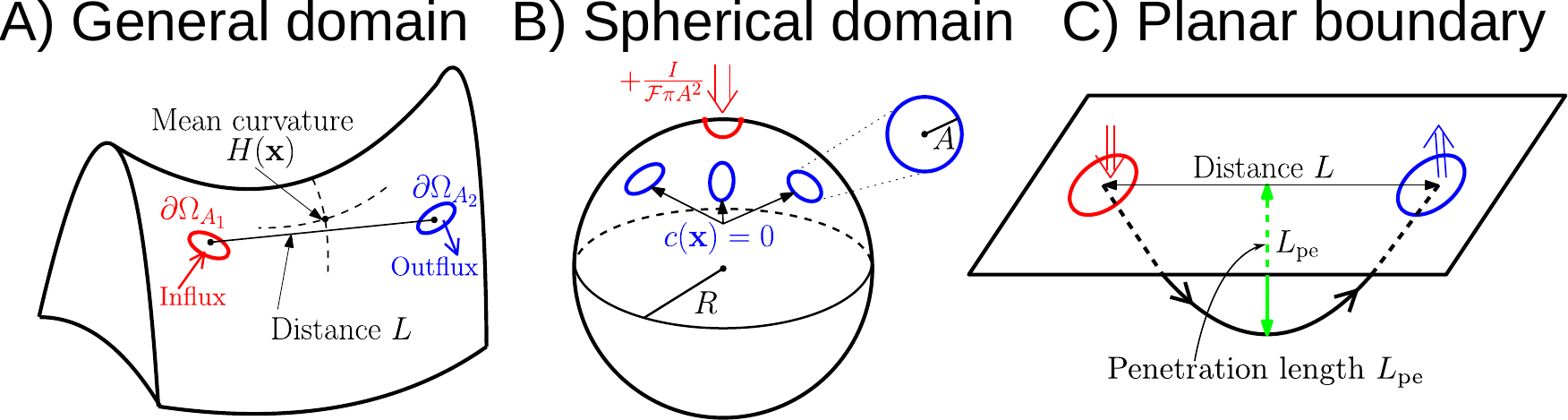}
\caption{\label{fig:fig1} \textbf{3-D geometrical domain types.} \textbf{A}) An arbitrary domain with general mean mean curvature function $H(\x)$. \textbf{B}) A ball domain of radius $R$. \textbf{C}) For two nearby narrow windows we can approximate the boundary by a tangent plane.}
\end{figure}

How far inside a domain an influx of Brownian particles entering through a narrow window can perturb the steady-state bulk concentration when they can escape through a neighboring window? We summarize in this section the properties of a diffusion system with one species inside a bounded domain between two narrow windows (Fig.~\ref{fig:fig1}\textbf{A})). This perturbation of concentration is obtained using asymptotic formulas. These formulas 
link together parameters and mix current and geometrical parameters (part \S \ref{sec:model_neumann}). In part \ref{sec:Lpe}, we recall how to estimate the size of nanodomains, where the concentration is spatially perturbed. A measure of the perturbation is computed using the length of penetration (introduced in \cite{paquin2021}), computed as the maximum distance from the boundary of the flow line going from the center of the influx window where particles are injected to the center of the neighboring window where they are expelled (Fig.~\ref{fig:fig1}\textbf{C})). Finally, we will see how a collection of exit windows (part \S \ref{sec:model_absorbing}) with Dirichlet absorbing boundary conditions (Fig.~\ref{fig:fig1}\textbf{B})) can affect the concentration distribution.
\subsection{Entry and exit of particles mediated by Neumann flux boundary conditions}\label{sec:model_neumann}
The geometry to consider is a bounded domain $\Omega$ with two small patches on the boundary $\p\Omega$. The two patches $\p\Omega_{A_i}$ for $i=1,\,2$ correspond to narrow circular windows of identical radius $A$, each centered at $\x_i$. The steady-state distribution of the concentration  $c(\x)$ for a single species diffusing (with diffusion coefficient $D$) inside $\Omega$ is influenced by the Neumann flux boundary conditions modeling the injection of a particles current of amplitude $I$ through the entry point $\p\Omega_{A_1}$, while the particles exit through $\p\Omega_{A_2}$ with current of opposite sign $-I$. The remaining boundary part $\p\Omega_r$ is reflecting. Furthermore, the domain size is such that the isoperimetric ratio is of order $O(1)$. The steady-state density $c(\x)$ is thus solution of the Laplace's equation
\begin{subequations}\label{eq:model_neumann}
\begin{align}
D\Delta c(\x) = 0\,, \qquad \x \in \Omega\,,
\end{align}
with boundary conditions
\begin{align}
D\frac{\p c(\x)}{\p\n} = \frac{I}{\FF \pi A^2}\,, \quad \x \in \p\Omega_{A_1}\,, \quad D\frac{\p c(\x)}{\p\n} = -\frac{I}{\FF \pi A^2}\,, \quad \x \in \p\Omega_{A_2}\,, \quad  \frac{\p c(\x)}{\p\n} = 0\,, \quad \x \in \p\Omega_r\,,
\end{align}
where $\FF$ is the Planck's constant. We fix the total concentration by choosing the constant $C_0$ as the mean density:
\beq
\frac{1}{|\Omega|}\int_\Omega c(\x)d\x = C_0\,.
\eeq
\end{subequations}
The question at hand is to study the difference of concentration $c(\x_1) - c(\x_2)$ with respect to the model parameters and the local window organization. For an almost convex or circular domain, we consider that the local radius of curvature $R$ of the boundary where are located the two windows is such that $R\approx\sqrt{|\p\Omega|}$. The windows are much smaller than the total boundary area, with $\sqrt{|\p\Omega_{A_i}|} \ll \sqrt{|\p\Omega|}$ yielding $A \ll R$. In that case, the concentration difference depends on the current $I$ and various geometrical parameters \cite{paquin2021}, and is given by
\beq\label{eq:cdiff_neumann}
c(\x_1) - c(\x_2) = \frac{2I}{\FF \pi A D}\left(1 - \frac{H(\x_1) + H(\x_2)}{8}A\log\left(\frac{A}{R}\right) + O\left(\frac{A}{R}\right)\right)\,,
\eeq
where $H(\x_i)$ is the mean curvature computed at the center $\x_i$ of the narrow windows \cite{paquin2021}. At leading-order in $A/R$, the difference increases linearly with the current amplitude $I$, while it is inversely proportional to the diffusion coefficient $D$ and the window radius $A$. Another key parameter is the Euclidean distance $L = \|\x_1 - \x_2\|$ between the two window centers, but its influence is restricted to higher-order terms of the expansion \eqref{eq:cdiff_neumann}. This expression shares similarity with the classical narrow escape theory \cite{HolcmanSchuss2015} with multiple targets, where the mean escape time is only affected by the targets geometrical organization via third-order terms \cite{cheviakov2010}.\\
A brief derivation of formula \eqref{eq:cdiff_neumann} uses the Neumann Green's function $G_s(\x;\y)$ with singularity $\y$ on the boundary $\p\Omega$. This Green's function is solution of the boundary value problem
\beq\label{eq:gf}
D\Delta G_s(\x;\y) = \frac{R}{|\Omega|}\,, \quad \x \in \Omega\,, \y \in \p\Omega\,, \quad D\frac{\p G_s(\x;\y)}{\p\n} = R\delta(\x-\y)\,, \quad \x,\y \in \p\Omega\,,
\eeq
with $\int_\Omega G_s(\x;\y)d\x = 0$, and its behavior near the singular diagonal $\x = \y$ is given explicitely by \cite{HolcmanSchuss2015}
\beq\label{eq:sing_behavior}
G_s(\x;\y) \approx \frac{R}{D}\left(\frac{1}{2\pi\|\x-\y\|} - \frac{H(\y)}{4\pi}\log\left(\frac{\|\x-\y\|}{R} \right) + O(1)\right)\,,
\eeq
where $H(\y)$ is the mean surface curvature in $\y$ on the boundary. The Green's identity given by
\beq\label{eq:greenid}
\int_{\Omega} D\left( G_s(\x;\y)\Delta c(\x) - c(\x) \Delta G_s(\x;\y) \right)d\x = \int_{\p \Omega} D\left( G_s(\x;\y)\frac{\p c(\x)}{\p \n} - u \frac{\p G_s(\x;\y)}{\p \n} \right)d\x\,,
\eeq
is used for the substitution of \eqref{eq:model_neumann} and \eqref{eq:gf} yielding
\beq
c(\y) = \frac{1}{|\Omega|}\int_{\Omega} c(\x) d\x + \frac{I}{\FF\pi A^2 R}\left(\int_{\p\Omega_{A_1}}G_s(\x;\y)d\x - \int_{\p\Omega_{A_2}}G_s(\x;\y)d\x\right)\,,
\eeq
thus the density difference between the two narrow windows is given by
\begin{equation}\label{eq:cdiff_temp}
\begin{split}
c(\x_1) - c(\x_2) = \frac{I}{\FF\pi A^2 R}&\left(\int_{\p\Omega_{A_1}}G_s(\x;\x_1)d\x + \int_{\p\Omega_{A_2}}G_s(\x;\x_2)d\x \right. \\
&\left. -\int_{\p\Omega_{A_2}}G_s(\x;\x_1)d\x - \int_{\p\Omega_{A_1}}G_s(\x;\x_2)d\x \right) \,.
\end{split}
\end{equation}
The integrals around the singularity of the Green's function are then evaluated directly using expansion \eqref{eq:sing_behavior} with radial coordinate $r = \|\x-\x_i\|$:
\begin{align}
\int_{\p\Omega_{A_i}}G_s(\x;\x_i)d\x &\approx \frac{2\pi R}{D}\int_0^A\left(\frac{1}{2\pi r} - \frac{H(\x_i)}{4\pi}\log\left(\frac{r}{R} \right) + O(1) \right)rdr\,, \nonumber \\
&\approx \frac{RA}{D}\left(1 -  \frac{H(\x_i)}{4}\log\left(\frac{A}{R}\right) + O\left(\frac{A}{R}\right)\right). \label{eq:sing_int}
\end{align}
The mixed integral terms can be incorporated within the remainder term since
\beq\label{eq:mixed_int}
\int_{\p\Omega_{A_i}}G_s(\x;\x_j)d\x \approx \pi A^2 G_s(\x_i; \x_j)\,.
\eeq
Hence by substituting \eqref{eq:sing_int} and \eqref{eq:mixed_int} within \eqref{eq:cdiff_temp} we recover the two-term asymptotic formula given above in eq.~\eqref{eq:cdiff_neumann}.
\subsubsection{Voltage-Current distribution in a spherical domain}
When the domain $\Omega$ consists of a ball of radius $R$, we can extend expansion \eqref{eq:cdiff_neumann} with an additional term involving the distance parameter $L=\|\x_1 - \x_2\|$. By using the mean curvature function $H(\x) = 1/R$ and the exact solution for the Green's function $G_s(\x;\y)$ on the sphere given by \cite{cheviakov2010}
\beq\label{eq:gf_sphere}
G_s(\x;\y) = \frac{R}{D}\left(\frac{1}{2\pi\|\x-\y\|} - \frac{1}{4\pi R}\log\left(\frac{\|\x-\y\|^2}{2R^2}+\frac{\|\x-\y\|}{R}\right) + \frac{1}{4\pi R}\left(\log(2) - \frac{9}{5}\right)\right)\,,
\eeq
with $\|\x\| = \|\y\| = R$ and $\x \neq \y$, we obtain the three-term asymptotic expansion \cite{paquin2021}
\beq\label{eq:threeTerm_neumann}
c(\x_1) - c(\x_2) = \frac{2I}{\FF\pi AD}\left(1 - \frac{A}{4R}\log\left(\frac{A}{R}\right) + \left( \frac{1}{8} - \frac{R}{2L} + \frac{1}{4}\log\left(\frac{L^2}{2R^2} + \frac{L}{R}\right)\right)\frac{A}{R} + O\left(\left(\frac{A}{R}\right)^2\right) \right)\,.
\eeq
Numerical simulations are used to explore how geometrical parameters can affect the concentration differences due to influx diffusion (Fig.~\ref{fig:fig2}). We solve \eqref{eq:model_neumann} using COMSOL \cite{comsol} on a spherical ball domain (see Appendix \ref{sec:mesh} for a mesh decomposition description). As predicted by eq.~\eqref{eq:threeTerm_neumann}, the density drop $c(\x_1) - c(\x_2)$ increases linearly with the current $I$ (Fig.~\ref{fig:fig2}\textbf{A})) and behaves as the reciprocal of the window radius $A$ (Fig.~\ref{fig:fig2}\textbf{C})). Finally, larger distances $L$ between the narrow windows also lead to steady-state concentrations being more spatially heterogeneous (Fig.~\ref{fig:fig2}).
\begin{figure}[!ht]
\centering
\includegraphics[width=\linewidth]{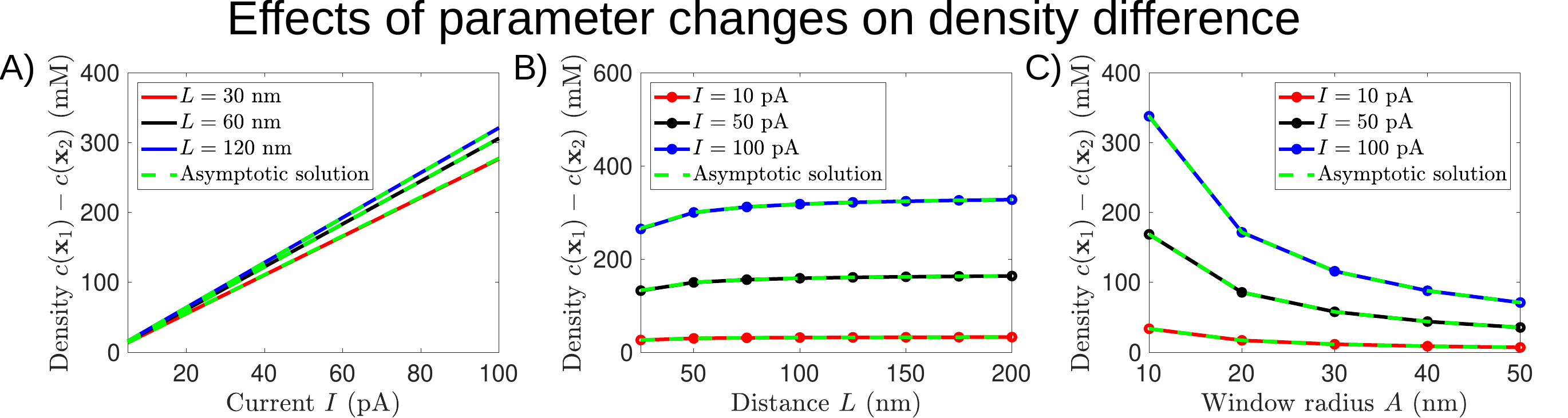}
\caption{\label{fig:fig2} \textbf{Spherical domain case.} Density difference between the influx and outflux as a function of the current amplitude $I$ (\textbf{A}), their distance $L$ from each other (\textbf{B}) and the narrow window radius $A$ (\textbf{C}). The green dashed curves indicate the asymptotic solution \eqref{eq:threeTerm_neumann}. Model parameters are $R = 500$ nm, $A = 10$ nm, $C_0 = 200$ mM, $D = 200\,\mu{\rm m}^2/s$. For the last case (\textbf{C}) the two windows are antipodes located, thus $L = 1\,\mu{\rm m}$. Values for other physical parameters are given in Table \ref{table:electrical_param}.}
\end{figure}
\subsection{Voltage-current relation in an infinite half-space and the notion of penetration length}\label{sec:Lpe}
\begin{figure}[!ht]
\centering
\includegraphics[width=0.75\linewidth]{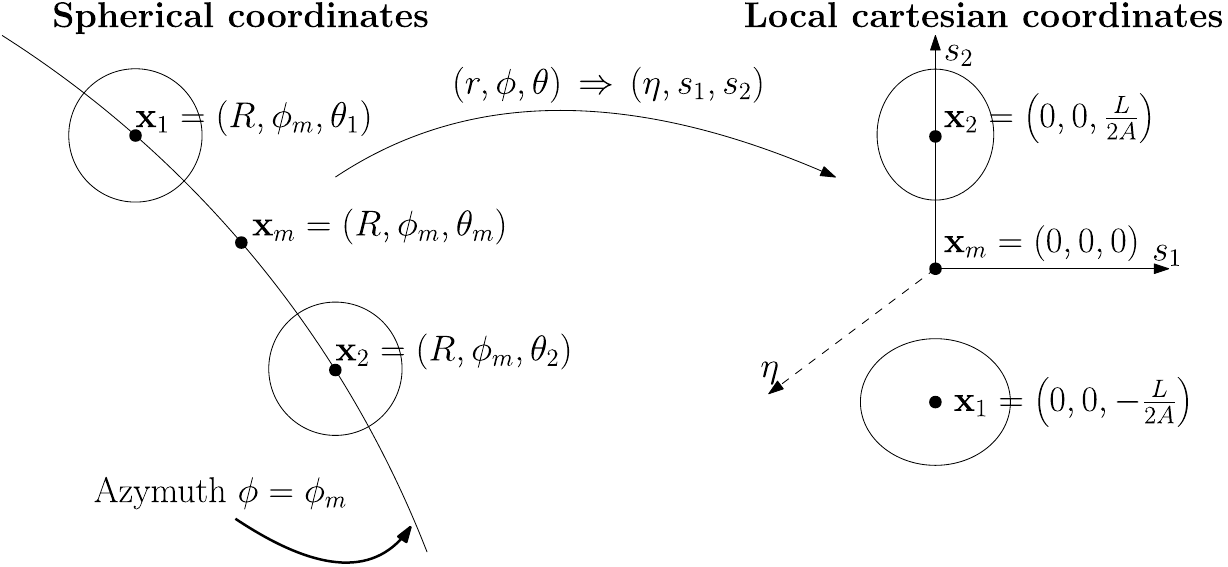}
\caption{\label{fig:local_cart_coord} \textbf{Projection of spherical to planar boundaries.} Using the coordinates \eqref{eq:cart_coord} we can map \eqref{eq:model_neumann} to a diffusion problem in the infinite half-space (eq.~\eqref{eq:inf_half}).}
\end{figure}

\begin{figure}[!ht]
\centering
\includegraphics[width=\linewidth]{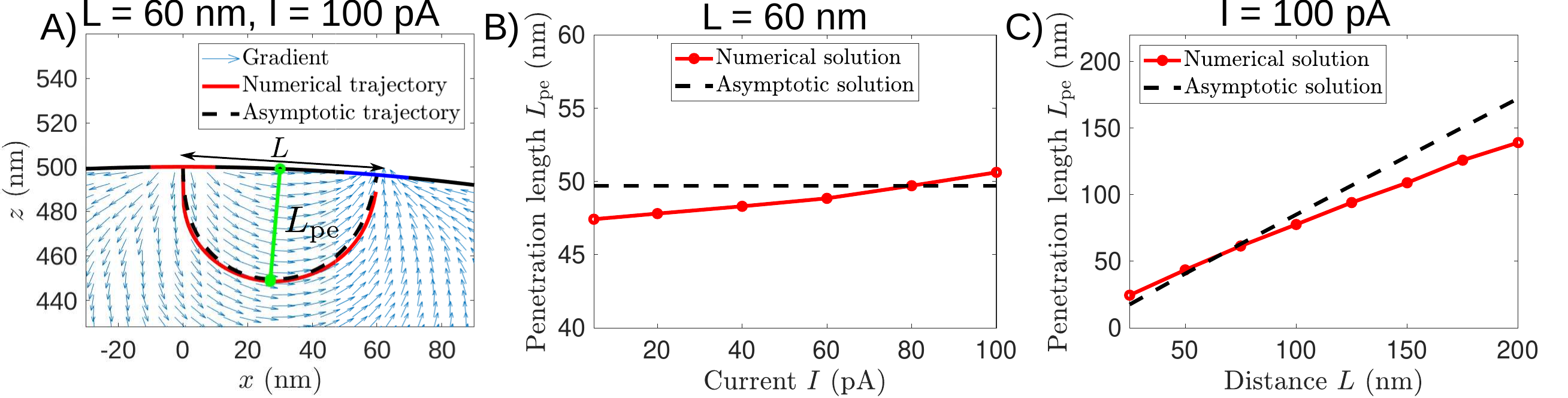}
\caption{\label{fig:fig3} \textbf{Field lines trajectories and penetration length.} (\textbf{A}) For a distance $L = 60$ nm and a current $I = 100$ pA, examples of trajectories where the gradient is computed using COMSOL \cite{comsol} (red curve) and from the asymptotic solution \eqref{eq:inner_sol} via the 2-D dynamical system \eqref{eq:dyna_neu} (black dashed curve). The green line indicates the penetration length $L_{\rm pe}$. (\textbf{B}) Penetration length $L_{\rm pe}$ versus the current amplitude $I$ (\textbf{C}) Penetration length $L_{\rm pe}$ versus the distance $L$ between the narrow windows. Parameter values are the same as in Fig.~\ref{fig:fig2}.}
\end{figure}

To measure how deep inside a medium can an influx current perturb the steady-state concentration, the penetration length $L_{\rm pe}$ was recently introduced based on model \eqref{eq:model_neumann}. The penetration length $L_{\rm pe}$ is  the maximal distance from the boundary achieved by trajectories following the gradient field lines from the influx current $I$ receiving to the exit windows.  Thus it is associated with paths of steepest concentration descent. Explicit solutions for the concentration gradient $\nabla c(\x)$ are in general not available for arbitrary domains $\Omega$. However, using a tangent plan approximation (on a smooth surface), the boundary layer analysis in the vicinity of two nearby (almost tangent) windows maps the solutions of \eqref{eq:model_neumann} into the ones of the Laplace's equation in the infinite half-space, with a planar boundary. This is illustrated below starting from a spherical ball domain.\\
The procedure is as follows: first, we use spherical coordinates $(r,\phi,\theta)$ to parameterize the narrow windows, with radius $r < R$, azymuth $\phi$ and colatitude $\theta$, yielding
\beq\label{eq:win_sphere}
\partial\Omega_{A_j} = \left\{ (R,\phi,\theta) \, \left| \, (\theta - \theta_j)^2 + \sin^2(\theta_j)(\phi - \phi_m)^2 \leq \left(\frac{A}{R}\right)^2 \right. \right\}\,, \quad j = 1,\,2\,.
\eeq
The two windows are aligned on the same azymuth $\phi_m$ and since they can be at most tangent the colatitudes satisfy $\theta_2 - \theta_1 > 2A/R$ . By then defining $\theta_m = (\theta_1 + \theta_2)/2$ as the midpoint colatitude coordinate, we can introduce a set of local cartesian coordinates $(\eta,s_1,s_2)$ \cite{cheviakov2010},
\beq\label{eq:cart_coord}
\eta = \frac{R-r}{A}\,, \quad s_1 = \frac{R}{A}\sin(\theta_m)(\phi - \phi_m)\,, \quad s_2 = \frac{R}{A}(\theta - \theta_m)\,,
\eeq
with $0 \leq \eta < \infty$ and $-\infty < s_1,s_2 < \infty$. With these new coordinates, we can map the narrow windows defined in eq.~\eqref{eq:win_sphere} onto the plane $\eta = 0$ (Fig.~\ref{fig:local_cart_coord}),
\beq
\partial\Omega_{A_j} = \left\{ (0,s_1,s_2) \, \left| \, \rho_i^2 \leq 1 \right. \right\}\,, \quad j = 1,\,2\,, \quad \text{with} \quad \rho_j = \sqrt{\left(\frac{\sin(\theta_j)}{\sin(\theta_m)}s_1\right)^2 + \left(s_2 + (-1)^{j+1} \frac{L}{2A}\right)^2}\,.
\eeq
The inner density variable is defined by
\beq\label{eq:inner_density}
\CC(\eta,s_1,s_2) = \frac{1}{C_0}c\left(R - \eta A,\phi_m + \frac{A}{R\sin(\theta_m)}s_1,\theta_m + \frac{A}{R}s_2\right)\,,
\eeq
and by substituting within the model \eqref{eq:model_neumann} we find that at leading-order (in terms of the ratio $A/R \ll 1$) it is solution in the infinite half-space of the Laplace's equation
\begin{subequations}\label{eq:inf_half}
\begin{align}
\frac{\p^2 }{\p \eta^2}\CC(\eta,s_1,s_2) + \frac{\p^2}{\p s_1^2}\CC(\eta,s_1,s_2) + \frac{\p^2}{\p s_2^2} \CC(\eta,s_1,s_2) = 0\,, \quad \eta > 0, \quad - \infty < s_1, s_2 < \infty.
\end{align}
Furthermore, the corresponding boundary conditions on the plane $\eta = 0$ are
\begin{align}\label{eq:inner_bc}
-\left.\frac{\p \CC}{\p\eta}\right|_{\eta = 0} &= \frac{I}{\FF \pi AC_0D}\,, \quad \rho_1 \leq 1 \quad
-\left.\frac{\p \CC}{\p\eta}\right|_{\eta = 0} = -\frac{I}{\FF \pi AC_0D}\,, \quad \rho_2 \leq 1\,, \\
\left.\frac{\p \CC}{\p\eta}\right|_{\eta = 0} &= 0\,, \quad \rho_1 > 1 \text{ and } \rho_2 > 1\,,
\end{align}
\end{subequations}
and the far-field condition is $\CC(\eta,s_1,s_2) \to 1$ as $\sqrt{\eta^2 + s_1^2 + s_2^2} \to \infty$. The solution of this boundary value problem (eq.~\eqref{eq:inf_half}) has an integral form \cite{carslaw1988},
\beq\label{eq:inner_sol}
\CC(\eta,s_1,s_2) = 1 + \frac{I}{\FF \pi AC_0D}\int_0^\infty e^{-m\eta}\frac{J_1(m)}{m} & \left(J_0\left( m \rho_1 \right) - J_0\left(m \rho_2 \right) \right) dm\,,
\eeq
where $J_n(x)$ is the Bessel function of order $n$. When calculating the difference between the entry and exit points we obtain
\beq
c(\x_1) - c(\x_2) = C_0\CC\left(0,0,-\frac{L}{2A}\right) - C_0\CC\left(0,0,-\frac{L}{2A}\right) = \frac{2I}{\FF \pi AD}\left(1 - \int_0^\infty J_1(m)J_0\left(m\frac{L}{A}\right)dm\right)\,,
\eeq
and interestingly for well-spaced windows, i.e.~with $L/A \gg 1$, the integral term vanishes and we recover the leading-order term from eq.~\eqref{eq:cdiff_neumann}. Finally,  from the time-dependent trajectories $C(t) = (\eta(t),s_1(t),s_2(t))^T$ solutions of the dynamical system
\beq\label{eq:dyna}
\frac{d}{dt}C(t)
= -\begin{pmatrix} \frac{\p}{\p \eta}\CC\left(\eta(t),s_1(t),s_2(t)\right) \\
\frac{\p}{\p s_1}\CC\left(\eta(t),s_1(t),s_2(t)\right) \\
\frac{\p}{\p s_2}\CC\left(\eta(t),s_1(t),s_2(t)\right)
\end{pmatrix}\,, \quad
C(0) = \begin{pmatrix} 0 \\ 0 \\ -\frac{L}{2A} \end{pmatrix}\,,
\eeq
with starting point at the center of the influx receiving window, we define the penetration length as $L_{\rm pe} = A\eta_{\rm pe}$, with $\eta_{\rm pe}$ the maximal distance from the planar boundary when following the trajectory $C(t)$,
\beq\label{eq:def1}
\eta_{\rm pe} = \max\left\{ \eta(t) \, \left| \, \frac{d}{dt}C(t) = -\nabla_{(\eta,s_1,s_2)} \CC \left(C(t)\right) \quad \text{with} \quad C(0) = (0,0,-L/(2A))^T \right. \right\}\,.
\eeq
Because of symmetry we can reduce \eqref{eq:dyna} to a 2-D dynamical system on the half-plane $s_1 = 0$ and $\eta > 0$, that is given by
\beq\label{eq:dyna_neu}
\frac{d}{dt}
\begin{pmatrix}
\eta \\
s_2
\end{pmatrix}
=
\frac{I}{\FF \pi AC_0D}
\begin{pmatrix}
\int_0^\infty e^{-m\eta}J_1(m)\left(J_0\left(m\left(\frac{L}{2A}+s_2\right)\right) - J_0\left(m\left(\frac{L}{2A}-s_2\right)\right)\right)dm \\
\int_0^\infty e^{-m\eta}J_1(m)\left(J_1\left(m\left(\frac{L}{2A}+s_2\right)\right) + J_1\left(m\left(\frac{L}{2A}-s_2\right)\right)\right)dm
\end{pmatrix}\,,
\eeq
with starting point
\beq
\begin{pmatrix}
\eta(0) \\
s_2(0)
\end{pmatrix} =
\begin{pmatrix}
0 \\
-\frac{L}{2A}
\end{pmatrix}\,.
\eeq
This reduced initial value problem can be solved solved numerically (Fig.~\ref{fig:fig3}). It allows to study how the current intensity $I$ and the inter-window distance $L$ affect the penetration length $L_{\rm pe}$. Interestingly, an influx current can penetrate on ranges between $50$ and $200$ nanometers, independent of the current amplitude (Fig.~\ref{fig:fig3}\textbf{B})) while being linearly proportional to the distance between the narrow windows (Fig.~\ref{fig:fig3}\textbf{C})). The discrepancy between the asymptotic and COMSOL numerical results, observed for large distance between the narrow windows, results from the membrane curvature. This aspect is neglected at leading-order within the boundary layer (eq.~\eqref{eq:inf_half}).
\subsection{Concentration distribution with multiple absorbing windows}\label{sec:model_absorbing}
An alternative model to \eqref{eq:model_neumann} consists of diffusing particles that are expelled from the domain through a collection of exit sites each with absorbing boundary conditions (Fig.~\ref{fig:fig1}\textbf{B})). The boundary $\p\Omega$ is now divided into a reflecting part $\p\Omega_r$, a single narrow influx window $\p\Omega_{A_1}$, and $N-1$ narrow absorbing windows $\p\Omega_{A_j}$ for $j=2,\ldots,N$. All the windows $\p\Omega_{A_j}$ are identical disks of radius A. By denoting the union of all exit sites as $\p\Omega_a = \cup_{j=2}^N \p\Omega_{A_j}$, the influx diffusion model \eqref{eq:model_neumann} is as follows
\begin{subequations}\label{eq:model_absorbing}
\begin{align}
D\Delta c(\x) = 0\,, \qquad \x \in \Omega\,,
\end{align}
with boundary conditions
\begin{align}
D\frac{\p c(\x)}{\p\n} = \frac{I}{\FF \pi A^2}\,, \quad \x \in \p\Omega_{A_1}\,, \quad c(\x) = 0\,, \quad \x \in \p\Omega_a\,, \quad  \frac{\p c(\x)}{\p\n} = 0\,, \quad \x \in \p\Omega_r\,.
\end{align}
\end{subequations}
Since $|\p\Omega_{A_j}| \ll |\p\Omega|$, the concentration $c(\x_1)$ at the center of the influx window
\cite{paquin2021} can be computed
\beq\label{eq:cdiff_absorbing}
c(\x_1) = \frac{I}{\FF\pi AD}\left(1 + \frac{\pi}{4(N-1)} - \left(H\left(\x_1\right) + \frac{1}{(N-1)^2}\sum_{i=2}^N H\left(\x_i\right) \right) \frac{A}{4}\log\left(\frac{A}{R}\right) + O\left(\frac{A}{R}\right) \right)\,,
\eeq
where  the mean curvature $H(\x_j)$ is computed at the center $\x_j$ of window $\p\Omega_{A_j}$. The derivation uses the classical Weber's solution \cite{crank1975} to approximate the outward flux at each exit site,
\beq\label{eq:weber}
D\frac{\p c(\x)}{\p \n} = \frac{K_j}{\sqrt{A^2 - \|\x-\x_j\|^2}}\,, \quad \text{for} \quad \|\x-\x_j\| < A \,, \quad j \neq 1\,,
\eeq
where the constant $K_j$ controls the strength of the exit flux through the boundary $\p\Omega_{A_j}$. By integrating \eqref{eq:weber} over each window area $\p\Omega_{A_j}$, we get that the total outflux $\Phi_j$, in units of ${\rm mol}/{\rm s}$, is given by
\beq
\Phi_j = \int_{\p\Omega_{A_j}} D\frac{\p c(\x)}{\p \n} d\x = 2\pi K_j \int_0^A \frac{rdr}{\sqrt{A^2 - r^2}} = 2\pi A K_j\,, \quad j \neq 1.
\eeq
The contribution from all exiting fluxes must compensate the total influx $\Phi_1 = I/\FF$ (compatibility condition) thus linking together the unknown constants $K_j$,
\beq
\sum_{j=2}^N K_j = -\frac{I}{\FF 2\pi A}\,.
\eeq
For a general domain $\Omega$,  this formula involves the mean curvature function $H(\x)$ \cite{paquin2021},
\beq\label{eq:Phi_j}
\Phi_j = - \frac{I}{(N-1)\FF}\left( 1 - \frac{A\log(A/R)}{\pi(N-1)}\sum_{\substack{i=2\\ i\neq j}}^N \left(H(\x_i) - H(\x_j)\right) + O\left(\frac{A}{R}\right) \right)\,, \quad j \neq 1\,.
\eeq
\subsubsection{Explicit solution on a spherical domain}
\begin{figure}[!htb]
\centering
\includegraphics[width=\linewidth]{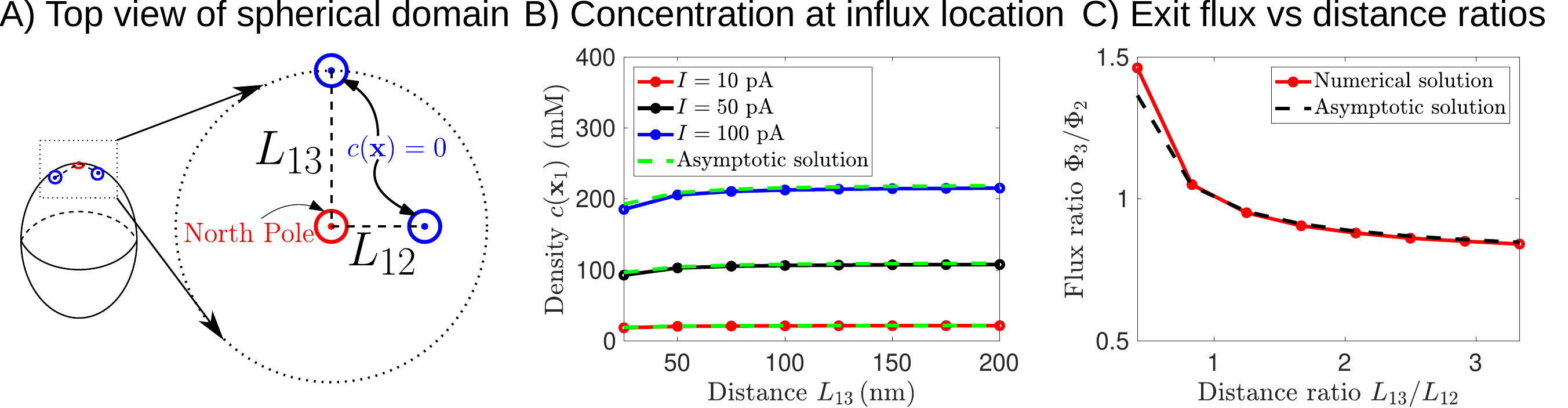}
\caption{\label{fig:fig4} \textbf{Splitting fluxes between two exit sites.} \textbf{A}) Top view of a spherical domain with $N=3$ narrow windows, with $L_{12}$ and $L_{13}$ denoting the distances between the absorbing and influx receiving windows. \textbf{B}) Density $c(\x_1)$ at location of  influx versus the distance $L_{13}$. The green dashed curves indicate the asymptotic solution \eqref{eq:cdiff_sphere_abs}. \textbf{C}) Ratio of exit fluxes versus ratio of distances, showing both COMSOL \cite{comsol} numerical (red curve) and asymptotic (eq.~\ref{eq:ratio_flux}, black dashed curve). The influx window is centered at the North Pole, while the fixed distance parameter is $L_{12} = 60$ nm. Other parameter values are the same as in Fig.~\ref{fig:fig2}.}
\end{figure}

For ball domains of radius $R$, since the exact Green's function \eqref{eq:gf_sphere} involves a three-term asymptotic expansion, the concentration has the following expression \cite{paquin2021}
\begin{equation}\label{eq:cdiff_sphere_abs}
\begin{split}
& c(\x_1) = \frac{I}{(N-1)\FF \pi AD}\left(N-1 + \frac{\pi}{4} - \frac{NA}{4R}\log\left(\frac{A}{R}\right) + \left( \frac{N+1}{8} - \frac{\log(2)}{4} \right. \right. \\
& \left. \left. + \frac{1}{N-1}\sum_{i=2}^N\sum_{j=i+1}^N \left( \frac{R}{L_{ij}} - \frac{1}{2}\log\left( \frac{L_{ij}^2}{2R^2} + \frac{L_{ij}}{R} \right) \right) - \sum_{j=2}^N \left( \frac{R}{L_{1j}} - \frac{1}{2}\log\left( \frac{L_{1j}^2}{2R^2} + \frac{L_{1j}}{R} \right) \right)\right)\frac{A}{R} + O\left(\frac{A^2}{R^2}\right)\right)\,,
\end{split}
\end{equation}
revealing the role of the windows geometrical organization on the maximal concentration drop across the domain. For a single exit window, this formula reduces to
\begin{equation}
\begin{split}
c(\x_1) &= \frac{I}{\FF \pi AD}\left(1+\frac{\pi}{4} - \frac{A}{2R}\log\left(\frac{A}{R}\right) \right. \\
& \left. + \left( \frac{3}{8} - \frac{\log(2)}{4} - \frac{R}{L} + \frac{1}{2}\log\left(\frac{L^2}{2R^2} + \frac{L}{R}\right) \right)\frac{A}{R} + O\left(\left(\frac{A}{R}\right)^2\right) \right)\,.
\end{split}
\end{equation}
Next, for the total outflux $\Phi_j$ through each exit window, the second term of the expansion \eqref{eq:Phi_j} vanishes since the mean curvature is constant on spherical domains. An additional term of order $O(A/R)$ yields  the formula below \cite{paquin2021}
\begin{equation}\label{eq:Phi_j_sphere}
\begin{split}
\Phi_j &= - \frac{I}{(N-1)\FF}\left(1 + \left(\sum_{\substack{i=2\\ i\neq j}}^N\left(\frac{R}{L_{1j}} - \frac{R}{L_{1i}} - \frac{R}{L_{ij}} - \frac{1}{2}\log\left( \frac{2R^2L_{1j}^2 + 4R^3L_{1j}}{\left(L_{1i}^2 + 2RL_{1i}\right)\left(L_{ij}^2 + 2RL_{ij}\right)} \right) \right) \right. \right. \\
& \left.\left. + \frac{2}{(N-1)} \sum_{i=2}^N \sum_{k=i+1}^N \left(\frac{R}{L_{ik}} - \frac{1}{2}\log\left( \frac{L_{ik}^2}{2R^2} + \frac{L_{ik}}{R} \right) \right)\right)\frac{2A}{\pi R} + O\left(\left(\frac{A}{R}\right)^2\right)\right)\,, \quad j \neq 1\,.
\end{split}
\end{equation}
Finally, numerical results illustrating how the relative position with respect to the influx window can affect the strength of the exit fluxes are shown in Fig.~\ref{fig:fig4}. In a spherical domain with three narrow windows (Fig.~\ref{fig:fig4}\textbf{A})), for which only the distance $L_{13}$ between the first (receiving the influx) and third windows is allowed to vary, eq.~\eqref{eq:Phi_j_sphere} with $N=3$ is used to compute the flux ratio $\Phi_3/\Phi_2$ between the third and second window,
\beq\label{eq:ratio_flux}
\frac{\Phi_3}{\Phi_2} = \frac{1 - \frac{2A}{\pi R}\left(\frac{R}{L_{12}} - \frac{R}{L_{13}} + \frac{1}{2}\log\left(\frac{L_{13}^2 + 2RL_{13}}{L_{12}^2 + 2RL_{12}}\right)\right) + O\left(\left(\frac{A}{R}\right)^2\right)}{1 + \frac{2A}{\pi R}\left(\frac{R}{L_{12}} - \frac{R}{L_{13}} + \frac{1}{2}\log\left(\frac{L_{13}^2 + 2RL_{13}}{L_{12}^2 + 2RL_{12}}\right)\right) + O\left(\left(\frac{A}{R}\right)^2\right)}.
\eeq
Comparing this against ratios numerically computed with COMSOL \cite{comsol}, we conclude that exit sites located near the influx receiving window contribute more to the escape dynamics than the farther ones (Fig.~\ref{fig:fig4}\textbf{C})).
\section{Concentration and voltage distributions in a two-charge electro-diffusion model}\label{sec:electro_two}
We consider here a two-species steady-state model with positive $c_p(\x)$ and negative $c_m(\x)$ charges, and voltage $v(\x)$. The two species are monovalent with diffusion coefficients $D_p$, $D_m$ respectively. We first consider an arbitrary domain $\Omega$ with mean membrane curvature function $H(\x)$ and characteristic length-scale $R$. The boundary $\p\Omega$ is smooth and is punctured by two narrow windows receiving and emitting an electrical current composed of positive charges only. The membrane is impermeable to negative charges. The positive and negative charge densities are each solution of the Nernst-Planck equations:
\begin{subequations}\label{eq:electro_two}
\begin{align}
\left\{
\begin{array}{ll}
D_p\left(\Delta c_p(\x) + \dfrac{e}{k_B \TT}\nabla (c_p(\x)\nabla v(\x))\right) &= 0\,, \quad \x \in \Omega\,, \\ &\\
D_m\left(\Delta c_m(\x) - \dfrac{e}{k_B \TT}\nabla (c_m(\x)\nabla v(\x))\right) & =0\,, \quad \x \in \Omega\,,
\end{array}
\right.
\end{align}
where $k_B\TT/e$ is the thermal voltage. The voltage $v(\x)$ is solution of the Poisson's equation, given by
\begin{align}\label{eq:ss_pois}
\Delta v(\x) = -\frac{\FF}{\eps\eps_0}\left(c_p(\x) - c_m(\x)\right) \,, \quad \x \in \Omega\,,
\end{align}
where $\eps$ and $\eps_0$ are the relative and vacuum permittivities (table \ref{table:electrical_param}). The boundary $\p\Omega$ is divided into a reflective part $\p\Omega_r$, and two narrow windows $\p\Omega_{A_1}$ and $\p\Omega_{A_2}$ receiving and emitting positive charge electrical currents. The windows are identical disks of radius $A$ centered in $\x_1$ and $\x_2$ on the boundary. For the positive species $c_p(\x)$, we  have
\begin{align}
& D_p\left(\frac{\p c_p}{\p \n} + \frac{e}{k_B \TT} c_p \frac{\p v}{\p \n}\right) = 0\,, \quad \x \in \p\Omega_r\,, \quad D_p\left(\frac{\p c_p}{\p \n} + \frac{e}{k_B \TT} c_p \frac{\p v}{\p \n}\right) = \frac{I}{\FF \pi A^2}\,, \quad \x \in \p\Omega_{A_1}\,, \\
& D_p\left(\frac{\p c_p}{\p \n} + \frac{e}{k_B \TT} c_p \frac{\p v}{\p \n}\right) = -\frac{I}{\FF \pi A^2}\,, \quad \x \in \p\Omega_{A_2}\,,
\end{align}
while no-flux condition holds everywhere on the boundary for the negative charge,
\begin{align}
D_m\left(\frac{\p c_m}{\p \n} - \frac{e}{k_B \TT} c_m \frac{\p v}{\p \n}\right) = 0\,, \quad \x \in \p\Omega\,.
\end{align}
Moreover global electro-neutrality, with average positive and negative densities equal to $C_0$, yields the constraints
\begin{align}
\frac{1}{|\Omega|}\int_\Omega c_p(\x) = C_0\,, \quad \frac{1}{|\Omega|}\int_\Omega c_m(\x) = C_0\,,
\end{align}
and therefore to satisfy the compatibility condition of the Poisson's equation (eq.~\ref{eq:ss_pois}), we impose that the total surface charge vanishes at the boundary
\begin{align}
\frac{\p v}{\p\n} = 0\,, \quad \x \in \p\Omega\,.
\end{align}
\end{subequations}
To compensate for the localized fluxes of positive charges and due to electrochemical forces, negative ion steady-state distribution must also become spatially heterogeneous. Thus, we can quantify the resulting density differences between the entry and exit points (see section \ref{sec:diffusion}), in relations with current and geometrical parameters. Moreover, asymptotic analysis allows to compute the density and voltage differences as
\begin{subequations}\label{eq:electro_gen}
\begin{align}\label{eq:cdiff_gen}
c_p(\x_1) - c_p(\x_2) = c_m(\x_1) - c_m(\x_2) = \frac{I}{\FF \pi AD_p}\left(1 - \frac{H(\x_1) + H(\x_2)}{8}A\log\left(\frac{A}{R}\right) + O\left(\frac{A}{R}\right)\right)\,,
\end{align}
while the voltage is the logarithm of the ratio between the entry $c_m(\x_1)$ and exit $c_m(\x_2)$ negative charge densities,
\begin{align}\label{eq:vdiff_gen}
v(\x_1) - v(\x_2) = \frac{k_B\TT}{e}\log\left(\frac{C_0 + \frac{I}{2\mathcal{F}\pi AD_p} \left(1 - \frac{H(\x_1) + H(\x_2)}{8}A\log\left(\frac{A}{R}\right) + O\left(\frac{A}{R}\right)\right)}{C_0 - \frac{I}{2\mathcal{F}\pi AD_p} \left(1 - \frac{H(\x_1) + H(\x_2)}{8}A\log\left(\frac{A}{R}\right) + O\left(\frac{A}{R}\right)\right)} \right)\,,
\end{align}
\end{subequations}
where $H(\x_j)$ is the mean curvature calculated at center $\x_j$, $j=1,2$. These formulas show that the membrane curvature can modulate the voltage \cite{cartailler2017jns,cartailler2019jmb,cartailler2018pccp}. Interestingly the deviation magnitudes predicted by formulas eq.~\eqref{eq:cdiff_gen} and eq.~\eqref{eq:cdiff_neumann} (diffusion only) relate by a factor one-half. To conclude less significant density differences are achieved with electro-diffusion than for diffusion.\\
When an absorbing condition $c_p(\x) = 0$ is imposed on the exit window $\p\Omega_{A_2}$, an alternative formula for the voltage drop is obtained by considering the following ansatz
\beq\label{eq:abs_asym}
v(\x_1) - v(\x_2) = \frac{k_B\TT}{e}\log\left(1 + \frac{A^{2\alpha-1}}{2\gamma R^\alpha}\left(\frac{I}{\FF D\pi C_0}\right)^{1-\alpha}F\left(\frac{A}{R}\right)\right)\,,
\eeq
where $F\left(\frac{A}{R}\right)$ is defined as
\beq \label{ffunction}
F\left(\frac{A}{R}\right) =
1 + \frac{\pi}{4} - \frac{H(\x_1) + H(\x_2)}{4}A\log\left(\frac{A}{R}\right) + O\left(\frac{A}{R}\right)\,.
\eeq
Here the exponent $\alpha$ and the parameter $\gamma$ are found by numerical fit.
\subsection{Electro-diffusion with two species in a spherical domain}
\begin{figure}[!ht]
\centering
\includegraphics[width=\linewidth]{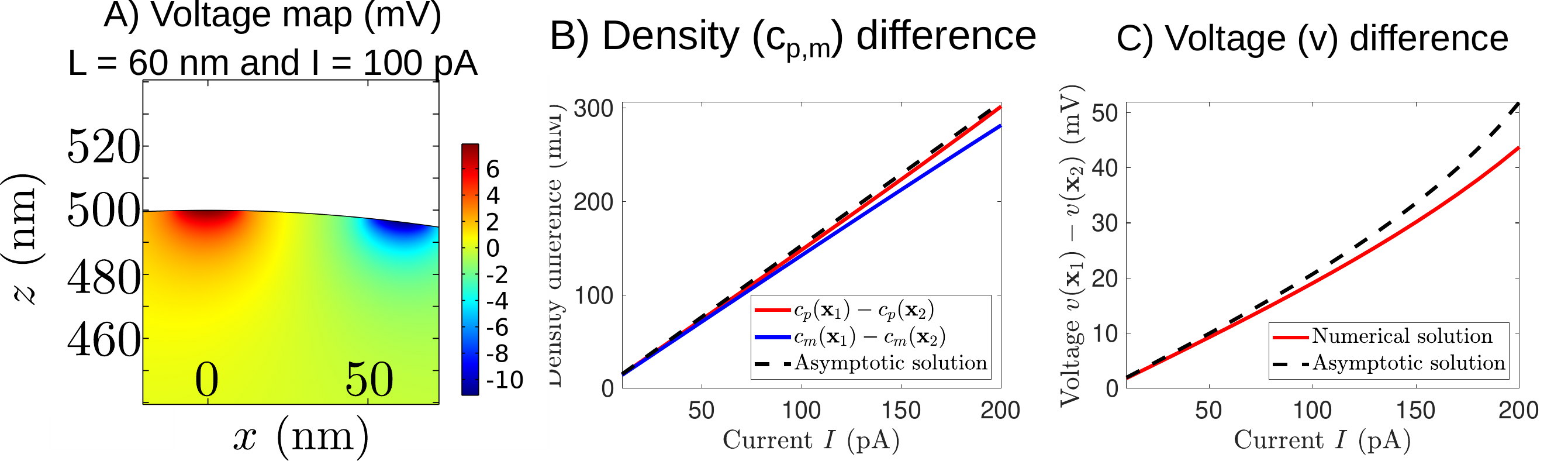}
\caption{\label{fig:fig5} \textbf{Nonlinear voltage-current relations.} \textbf{A}) 2-D voltage map obtained with COMSOL \cite{comsol}, with distance $L=60$ nm between the narrow windows and current $I=100$ pA. The influx window is located at the North Pole. \textbf{B}) Positive (red curve) and negative (blue curve) density differences versus injected current $I$. The black dashed-curve indicates the asymptotic solution \eqref{eq:electro_cdiff_sphere}. \textbf{C}) Potential drop versus injected current $I$, showing both COMSOL numerical (red curve) and asymptotic (eq.~\eqref{eq:vdiff_sphere}, black dashed curve) solutions. In (\textbf{B}-\textbf{C}) the distance is $L=60$ nm and other parameter values are taken from table \ref{table:electrical_param}.}
\end{figure}

For the case of a unit ball domain an additional term to the asymptotic formulas \eqref{eq:electro_gen} can be obtained, that reveals the role of the relative window positions on the voltage. The asymptotic expansion of $F\left(A/R\right)$ (eq.~\ref{ffunction}) leads to
\beq
F\left(\frac{A}{R}\right) =
1 - \frac{A}{4R}\log\left(\frac{A}{R}\right) + \left(\frac{1}{8} - \frac{R}{2L} + \frac{1}{4}\log\left( \frac{L^2}{2R^2} + \frac{L}{R}\right)\right)\frac{A}{R} + O\left(\left(\frac{A}{R}\right)^2 \right)\,,
\eeq
and the density differences are of the same order for both species,
\begin{subequations}\label{eq:electro_sphere}
\begin{align}\label{eq:electro_cdiff_sphere}
c_p(\x_1) - c_p(\x_2) = c_m(\x_1) - c_m(\x_2) = \frac{I}{\FF \pi AD_p}F\left(\frac{A}{R}\right)\,,
\end{align}
while the voltage is given by
\begin{align}\label{eq:vdiff_sphere}
v(\x_1) - v(\x_2) = \frac{k_B\TT}{e}\log\left(\frac{C_0 + \frac{I}{2\mathcal{F}\pi AD_p}F\left(\frac{A}{R}\right)}{C_0 - \frac{I}{2\mathcal{F}\pi AD_p}F\left(\frac{A}{R}\right)} \right)\,.
\end{align}
\end{subequations}
Formula \eqref{eq:vdiff_sphere} predicts nonlinear voltage effects deviating from Ohm's law at large current amplitude, with those nonlinear effects also observed numerically (Fig.~\ref{fig:fig5}\textbf{C})). Moreover, we report voltage spatial variations on ranges of size of $10$ to $25$ nanometers (Fig.~\ref{fig:fig5}\textbf{A})). This range is well-beyond the Debye length $\lambda_D$, obtained from classical electrolyte theory \cite{andelman}, which for two monovalent ions leads to $\lambda_D = \sqrt{\frac{\eps \eps_0 k_B\TT}{2e^2N_AC_0}} \approx 0.7$ nm (parameter values are from table \ref{table:electrical_param}).
\begin{table}[!ht]
\centering
\caption{\label{table:electrical_param} \textbf{Model parameters and physical constants}}
\begin{tabular}{|lcl|}
\hline
\textbf{Parameter} & \textbf{Symbol} & \textbf{Value} \\
\hline
Domain length-scale & $R$ & $500\,{\rm nm}$ \\
Window radius & $A$ & $10\,{\rm nm}$ \\
Window distance & $L$ & $[25,\,200]$ nm \\
Diffusion coefficient & $D_{p,m}$ & $200\,\mu{\rm m}^2{\rm s}^{-1}$ \\
Concentration & $C_0$ & $200\,{\rm mM}$ \\
Current amplitude & $I$ & $100\,{\rm pA}$ \\
Electronic charge & $e$ & $\approx1.60 \times 10^{-19}\,{\rm C}$ \\
Boltzmann constant & $k_B$ & $\approx1.38 \times 10^{-23}\,{\rm J}\,{\rm K}^{-1}$ \\
Temperature & $\mathcal{T}$ & $298\,{\rm K}$ \\
Avogadro number & $N_A$ & $\approx6.02 \times 10^{23}\,{\rm mol}^{-1}$ \\
Relative permittivity & $\eps$ & $78.4$ \\
Vacuum permittivity & $\eps_0$ & $\approx8.85\times 10^{-12}\,{\rm C}\,{\rm V}^{-1}\,{\rm m}^{-1}$ \\
Faraday's constant & $\FF$ & $\approx96485.33\,{\rm C}\,{\rm mol}^{-1}$ \\
Thermal voltage & $k_B\mathcal{T}/e$ & $\approx 25.68 \,{\rm mV}$ \\
\hline
\end{tabular}
\end{table}
\section{Conclusions and perspectives}\label{sec:conclusion}
In this chapter, we presented models and analyses of voltage dynamics and ions distribution after their entry through a channel, their motion inside the bulk of a cell, until their eventual extrusion by a neighboring pump. In the diffusion approximation with pumps or channels modeled as small windows, an influx of ions generates a local difference of concentration between the source and a neighboring target window. Obtaining a framework to compute local ionic concentration differences and electrical activity allows to quantify nanodomains where voltage distribution triggers the opening of neighboring channels \cite{zamponi2011}.\\
The mathematical analysis is based on the Laplace's equation, which neglects the electrochemical effects due to the local electric field (section \S \ref{sec:diffusion}). Asymptotic formulas for the concentration difference between two narrow windows, each with a nonzero Neumann flux boundary condition (\S \ref{sec:model_neumann}), reveal that the concentration difference depends at first order on the window size, the current intensity $I$ and the diffusion coefficient $D$, while at second order it also depends on the mean curvature computed at the center of each window. Finally the third order involves the geometrical organization of all the windows via the explicit solution of the Green's function, with such a solution available for spherical ball domains.  These computations differ from the classical narrow escape theory \cite{JSP2004,holcmanPNAS2007,benichou2008,pillay2010,cheviakov2010,zawada2013,HolcmanSchuss2015,grebenkov2016,grebenkov2019} describing the mean escape time of a stochastic particle initially distributed either at a point or uniformly inside a bounded domain, with the escape occurring through one of several narrow absorbing windows located on an otherwise reflective surface. The average time to escape, or Narrow Escape Time, is solution of the Poisson's equation (and not the Laplace's equation as in section \S \ref{sec:diffusion}) with mixed Dirichlet-Neumann boundary conditions, with small ratio of Dirichlet versus reflecting parts.\\
To measure how far from the channel the concentration is perturbed, we can use the penetration length. This length is constructed using a field lines approach, by measuring the maximal distance from the boundary for trajectories connecting the flux receiving and emitting windows (\S \ref{sec:Lpe}). The penetration length is independent of the current intensity $I$ of the field, but depends linearly on the distance $L$ between the windows. In \S \ref{sec:model_absorbing} we discussed multiple exit windows each with Dirichlet absorbing boundary conditions. Numerical and asymptotic results confirmed how the exit flux strength is mediated by the relative position with respect to the entry site.\\
Finally, the electric field generated by two charged ions was modeled by the electro-diffusion theory in section \S \ref{sec:electro_two}, where the localized flux current was composed of positive charges only. From density difference formulas we obtained voltage-current laws emphasizing the role of geometrical and membrane curvature on the voltage dynamics. We also reported interior voltage simulation results suggesting the formation of a nanodomain, with voltage effects extending well-beyond the Debye screening length from the classical electrolytes theory \cite{andelman}. To conclude, this present chapter summarized briefly modeling approaches, and relations between biophysical parameters and geometrical organization, that elucidate how concentration and voltage could change around voltage-gated channels, a key property for controlling the channel open probability by voltage in dendrites or small protrusions of neuronal cells \cite{holcman2015nrn}.
\section*{Acknowledgements}
F.P-L.~gratefully acknowledges the support from the Fondation ARC through a postdoctoral fellowship (award No ARCPDF12020020001505). D.H.~was supported by the European Research Council (ERC) under the European Union’s Horizon 2020 Research and Innovation Program (grant agreement No 882673), and ANR AstroXcite.
\begin{appendix}
\section{Appendix: Domain mesh and COMSOL simulations}\label{sec:mesh}
\begin{figure}[!ht]
\centering
\includegraphics[width=\textwidth]{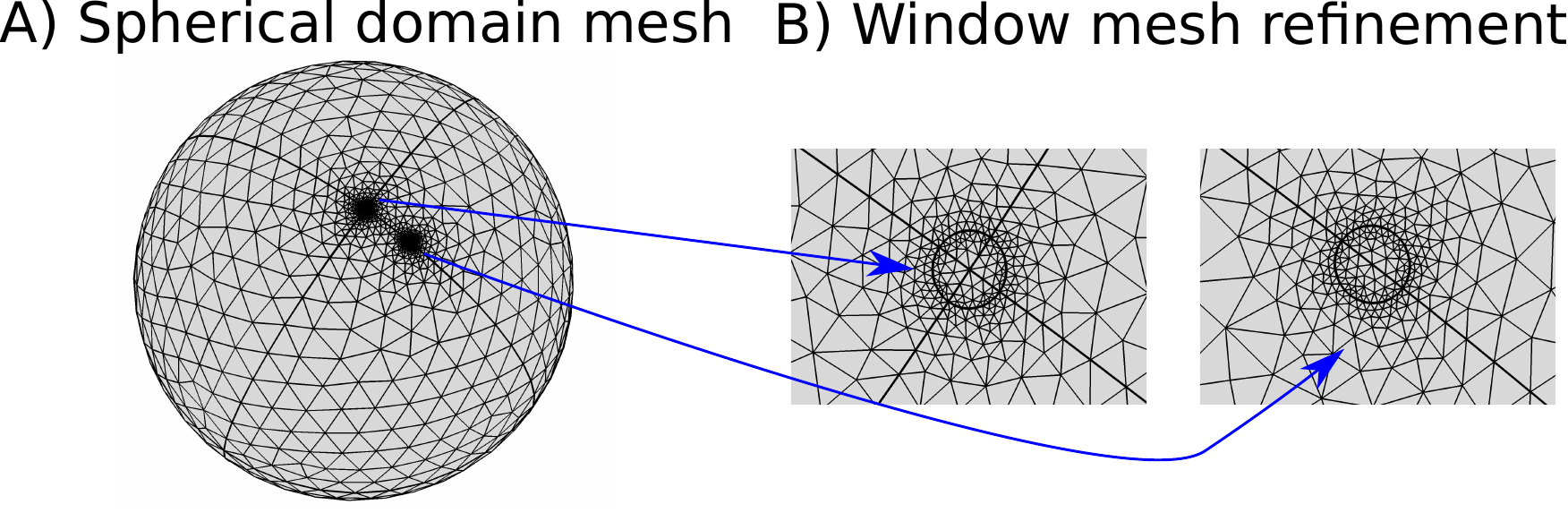}
\caption{\label{fig:mesh} \textbf{Spherical domain mesh.} \textbf{A}) The domain mesh is selected to be \textit{fine}. \textbf{B}) The narrow windows mesh is locally refined to be \textit{extremely fine}.}
\end{figure}

The \textit{Coefficient Form PDE} module from COMSOL Multiphysics version 5.2a \cite{comsol} is used to numerically solve the boundary value problems. All simulations are carried on 3-D spherical ball domains discretized with a free tetrahedral mesh that is \textit{extremely fine} on the narrow absorbing targets and \textit{fine} elsewhere on the domain (Fig.~\ref{fig:mesh}). The predefined mesh preferences from COMSOL, which control the maximum and minimum element sizes, vary from \textit{extremely coarse} to \textit{extremely fine}.
\end{appendix}
\normalem
\bibliographystyle{ieeetr}
\bibliography{ref}
\end{document}